# Sm$_2$Al: Another ferromagnet with spin-orbital compensation?

Pramod Kumar†, K. G. Suresh† and A. K. Nigam‡

†Department of Physics, I.I.T. Bombay, Mumbai 400076, India

‡Tata Institute of Fundamental Research, Homi Bhabha Road, Mumbai 400005, India

*Abstract*

*We report the observation of a compensation point in the temperature dependence of magnetization data of polycrystalline Sm$_2$Al. Magnetization measurements show that this compound magnetically orders at about 150 K. Below this temperature, the magnetization data shows a compensation point, which shifts with field. Hysteresis loops obtained below the compensation temperature show that the compound possesses exchange anisotropy. Both the exchange anisotropy field and the coercive field are found to be quite large and comparable to those of the classical spin-orbit compensated ferromagnet (Sm,Gd)Al$_2$. The heat capacity data also suggest that there are similarities between (Sm,Gd)Al$_2$ and Sm$_2$Al.*

Ever since the discovery[1-4] of the spin and orbital moment compensation in SmAl$_2$, there has been a lot of interest in identifying other Sm-based materials which show a similar phenomenon. A ferromagnet with no net magnetization is of significant interest, both from the point of view of applications and fundamental interest. Though anomalous magnetic properties are seen to be quite common in many Sm-containing materials, the absence of net magnetization is not very prevalent. Among the magnetic lanthanide ions, $Sm^{3+}$ is different in the sense that the crystal field splitting between the ground state and a few higher multiplets is not large and therefore, it results in considerable mixing of the ground and excited states. Another peculiarity with $Sm^{3+}$ is that the magnitudes of spin and orbital moments are nearly equal (~4 $\mu_B$). This, together with the fact that the temperature dependencies of these two contributions are not the same, leads to compensation point below the ordering temperature. By manipulating the relative magnitudes of spin and orbital magnetic moments, the delicate balance can be adjusted to result in the tuning of the compensation point. However, to the best of our knowledge, other than (Sm,Gd)$Al_2$, there are no other systems in which such an observation has been experimentally reported. As part of our study on the intermetallic series, we recently studied $R_2Al$ compounds with various rare earths. As expected, we find that the behavior of $Sm_2Al$ is quite different as compared to all other members of the series. On further analysis of the results on this compound, interestingly, we find that this system also shows a compensation point, resulting in zero net magnetization in the ordered state. This is presumably the first report on a pure Sm compound exhibiting this behavior. The results obtained in this compound are presented in this paper. Though the crystal structure is the same for other compounds of this series, the anomalous behavior presented in this paper has been observed only in $Sm_2Al$.

Polycrystalline sample of Sm$_2$Al was synthesized by conventional methods of arc melting and annealing. The annealed sample was characterized by room temperature powder x-ray diffractograms (XRD), collected using Cu-K$_\alpha$ radiation. The compositional analysis was carried out using EDAX (Energy Dispersive Analysis of X-rays) facility attached to SEM (Scanning Electron Microscope). The magnetization (M) has been measured under 'zero-field-cooled' (ZFC), 'field-cooled warming' (FCW) and 'field-cooled cooling' (FCC) conditions, in the temperature (T) range of 5-300 K and upto a field (H) of 50 kOe. The heat capacity (C) measurement has been performed in the temperature range 2-300 K up to a field of 90 kOe.

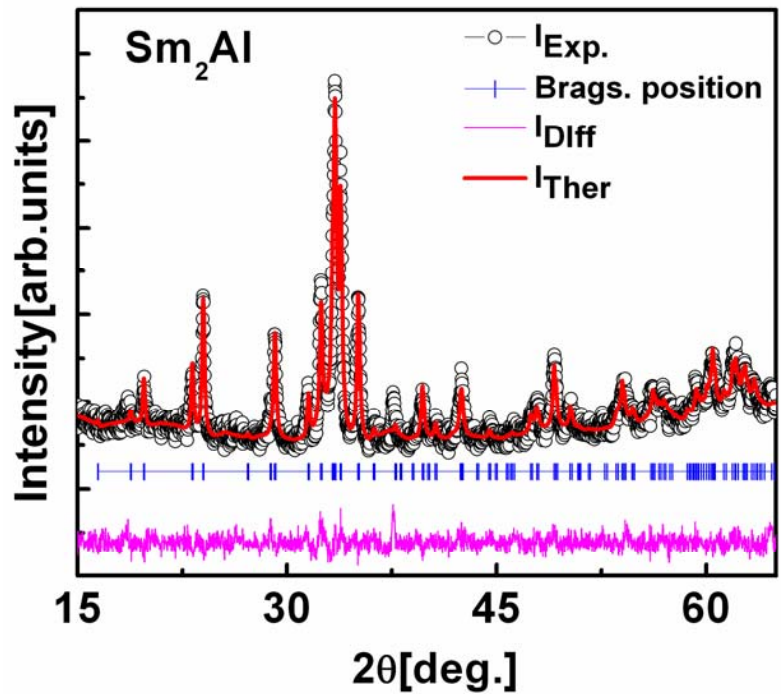

**Figure 1** Rietveld fitted powder x-ay diffraction pattern of Sm$_2$Al. Open circles show the experimental data and the solid red lines show the theoretical plot. The plot at the bottom shows the difference between the experimental and theoretical data.

Fig.1 shows the Rietveld refined powder x-ray diffractogram of $Sm_2Al$ taken at room temperature. As can be seen from this figure, the compound has formed in single phase in the orthorhombic structure (space group: Pnma) with no detectable impurity phases. The phase purity of the sample was also confirmed by the EDAX carried out on SEM micrograph. The lattice parameters obtained from the Rietveld refinement are $a$=6.633(1) Å, $b$=5.191(1) Å and $c$=9.580(1) Å. These values are in close agreement with those reported earlier[5].

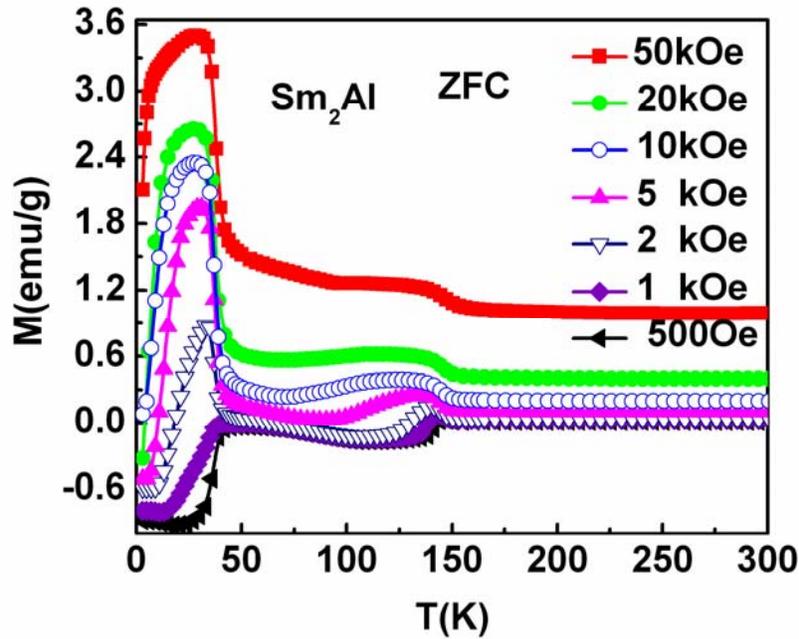

**Figure 2** Temperature variation of ZFC magnetization in $Sm_2Al$ in different fields.

Fig. 2 shows the ZFC magnetization of $Sm_2Al$ at various fields. It can be clearly seen from these plots that on decreasing the temperature, the magnetization shows an increase at about 150 K ($T_{ord}$). Below this temperature, the magnetization goes through a broad maximum followed by a minimum, with decrease in temperature. The temperature corresponding to the minimum is denoted as the compensation temperature ($T_{comp}$). For applied fields below 5 kOe, the minimum corresponds to negative values of magnetization. On further reducing the temperature, the magnetization turns positive and increases sharply for fields above 2 kOe, ultimately showing a maximum at about 40 K. For fields higher than 5 kOe, the trend of the ZFC magnetization is similar, except that the minimum is not well defined and that the minimum value is positive. At the lowest temperature, the magnetization is found to be negative for fields except in 50 kOe.

The magnetic ordering temperature, $T_{ord}$, calculated from the maximum in the $dM/dT$ vs. $T$ plots (ZFC) shows that it increases slightly with increase in the applied field, which suggests that the magnetic transition at this temperature is from paramagnetic to ferromagnetic, upon cooling. It can also be noted that the compensation temperature decreases with increase in field, except at 50 kOe. The value of $T_{comp}$ is found to be about 140K at 500 Oe, which decreases to 65 K as the field is increased to 20 kOe. At 50 kOe, $T_{comp}$ is about 90 K. The reduction in the magnetization below 40 K may be due to the domain wall pinning effect. However, at this point, we do not rule out some antiferromagnetic correlations below 40 K, which may also contribute to the thermomagnetic irreversibility. But, the high temperature (160-300K) susceptibility is found to show the Curie-Weiss behavior $\left[ \chi = \chi_0 + \dfrac{C}{(T - \theta_P)} \right]$ very well with a positive paramagnetic Curie temperature ($\theta_P$) of 140 K.

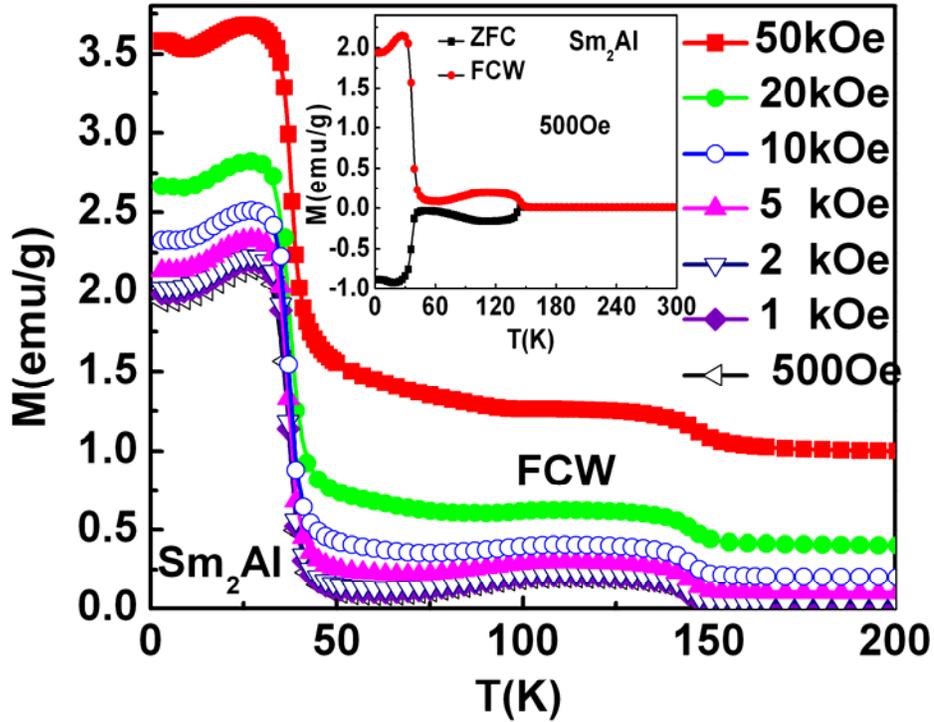

**Figure 3** Temperature variation of FCW magnetization in $Sm_2Al$ in different fields. The inset shows the ZFC and FCW magnetization data at 500 Oe.

Fig.3 show the FCW data of $Sm_2Al$, collected in various fields. It can be seen that the FCW magnetization behavior is nearly identical to the ZFC behavior. The main difference is that the magnetization is positive for all fields and at all temperatures, in this case. It is noteworthy that a reduction in the compensation temperature occurs in the FCW mode as compared to the ZFC mode, especially at low fields. For example, the $T_{comp}$ observed from the FCW data at 500 Oe is about 60 K, whereas it is about 90 K at 20 kOe. As in the case of ZFC data, $T_{ord}$ increases with field in the case of FCW data as well. Furthermore, the transition temperature (at about 40 K) and the sharpness of the this transition remain unchanged both in FCW and ZFC modes and for various applied fields. The inset of this figure shows the ZFC and FCW magnetization data in a field of 500 Oe. It was also observed that there is a small thermal hysteresis between the FCC and FCW data.

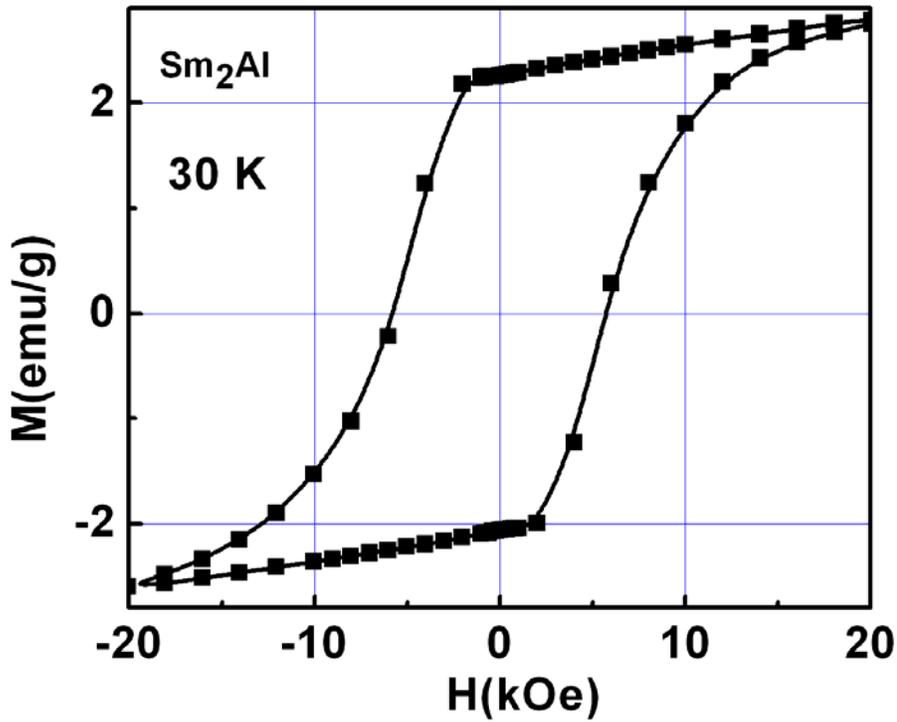

**Figure 4** Hysteresis plots of $Sm_2Al$ at 30 K

Fig. 4 shows the hysteresis loop at 30 K obtained after cooling the sample in a field of 10 kOe. The loop is asymmetric with respect to the magnetization axis, as seen in the case of materials with exchange bias. The exchange bias field has been calculated using the relation $H_{ex} = -\dfrac{(H_+ + H_-)}{2}$ and the coercivity has been calculated using the relation $H_C = \frac{(|H_+| + |H_-|)}{2}$. The hysteresis loop has been measured at 60 and 70 K also and the Table 1 gives the values of $H_{ex}$ and $H_C$ obtained at different temperatures. It is of interest to note that the maximum values of the corresponding parameters in the case of (Sm,Gd)Al$_2$ are ~600 Oe and 2 kOe, respectively[3].

**Table 1** Exchange anisotropy field and coercive field in Sm$_2$Al obtained at various temperatures.

| T (K) | $H_{ex}$ (Oe) | $H_C$ (kOe) |
| --- | --- | --- |
| 30 | 95 | 8.2 |
| 60 | 67 | 1.6 |
| 70 | 65 | 1.4 |

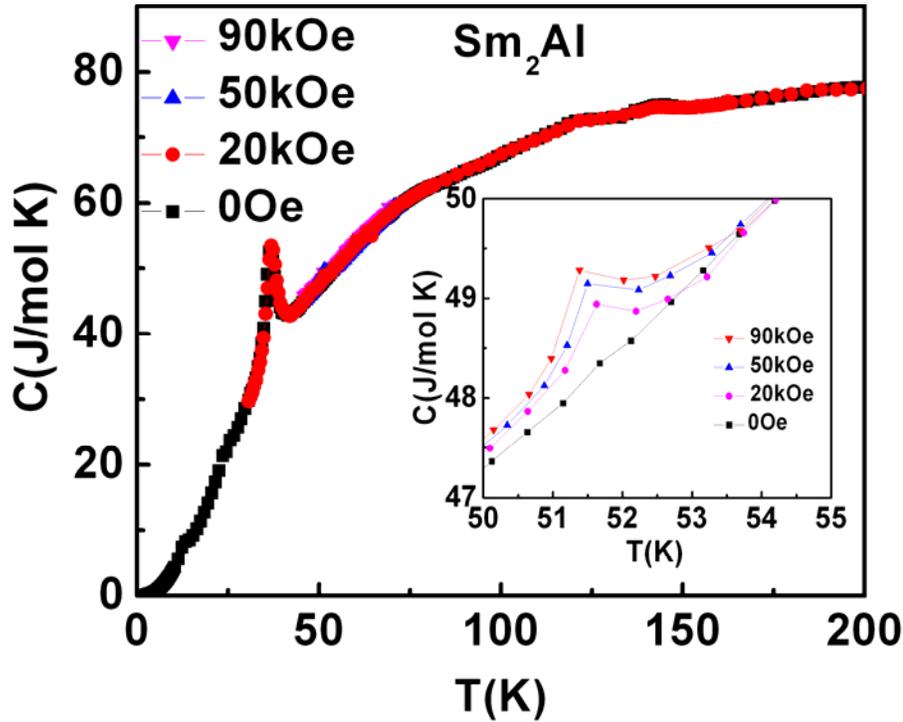

**Figure 5** Temperature variation of zero field heat capacity in Sm$_2$Al in various applied fields. The inset shows the expanded region of 50-55 K.

Fig. 5 shows the temperature variation of heat capacity in various applied fields. It shows a weak peak at about 150 K, but a prominent one at about 40 K, corresponding to the increase in the magnetization. From this figure, no anomaly could be seen at temperature corresponding to the compensation point. Therefore, the region around the compensation temperature has been expanded and is shown in the inset. It can be seen from the inset that there is an increase in the magnitude of heat capacity at temperatures close to 52 K. It may be noted that this temperature nearly coincides with the compensation temperature observed in the FCW M-T plots, especially at low fields. This observation is also similar to that reported in (Sm,Gd)Al$_2$, reported by Chen et al.[3].

From the results presented above, it is quite clear that the magnetic behavior of Sm$_2$Al is almost identical to that of the classical 'spin-orbital compensated ferromagnet' (Sm,Gd)Al$_2$. The observation of negative magnetization, compensation point, increase in the magnetization value at the compensation point with the applied field (conduction electron polarization effect) and the absence of a peak in the zero-field heat capacity at the compensation temperature observed in the present case show striking similarities with those observed in (Sm,Gd)Al$_2$. The observation of shifted hysteresis loops with exchange anisotropy fields and coercive fields comparable to (or larger than) those of (Sm,Gd)Al$_2$ is another striking similarity. However, there seems to be some differences between these two systems. The main difference is the magnetic behavior at very low temperatures. While in (Sm,Gd)Al$_2$, the magnetization continuously increases down to the lowest temperature, in the case of Sm$_2$Al, the magnetization goes through a maximum at about 40 K and then decreases. This presumably indicates some antiferromagnetic correlations at low temperatures. The low temperature magnetic state needs to be understood with additional measurements. As in the case of (Sm,Gd)Al$_2$, this may be due to the formation of two antiferromagnetically coupled sublattices at the molecular level[3]. It should be kept in mind that the crystal structure of (Sm,Gd)Al$_2$ is cubic while that of Sm$_2$Al is orthorhombic. There are two crystallographically inequivalent sites for the rare earth in the unit cell of R$_2$Al compounds. If the R moments on these two sites couple antiferromagnetically and if the temperature dependences of magnetization associated with the moments on these two sites are quite different, one can expect a compensation, as observed in many systems. However, no such compensation point has been seen in the M-T plots of any other member of the $R_2Al$ series, though they are iso-structural to $Sm_2Al$. In view of these, it is quite possible that spin-orbital moment compensation is the reason for the minimum in the magnetization and the other observed anomalies in $Sm_2Al$. In conclusion, we find that the magnetic and thermal properties of $Sm_2Al$ are more or less similar to those of the spin-orbital compensated $(Sm,Gd)Al_2$. From the present study, $Sm_2Al$ appears to be the first undoped system to exhibit spin-orbit compensation.

**Acknowledgement:**

The authors thank Pravin Patade and Devendra Buddikot for their help in the measurements.

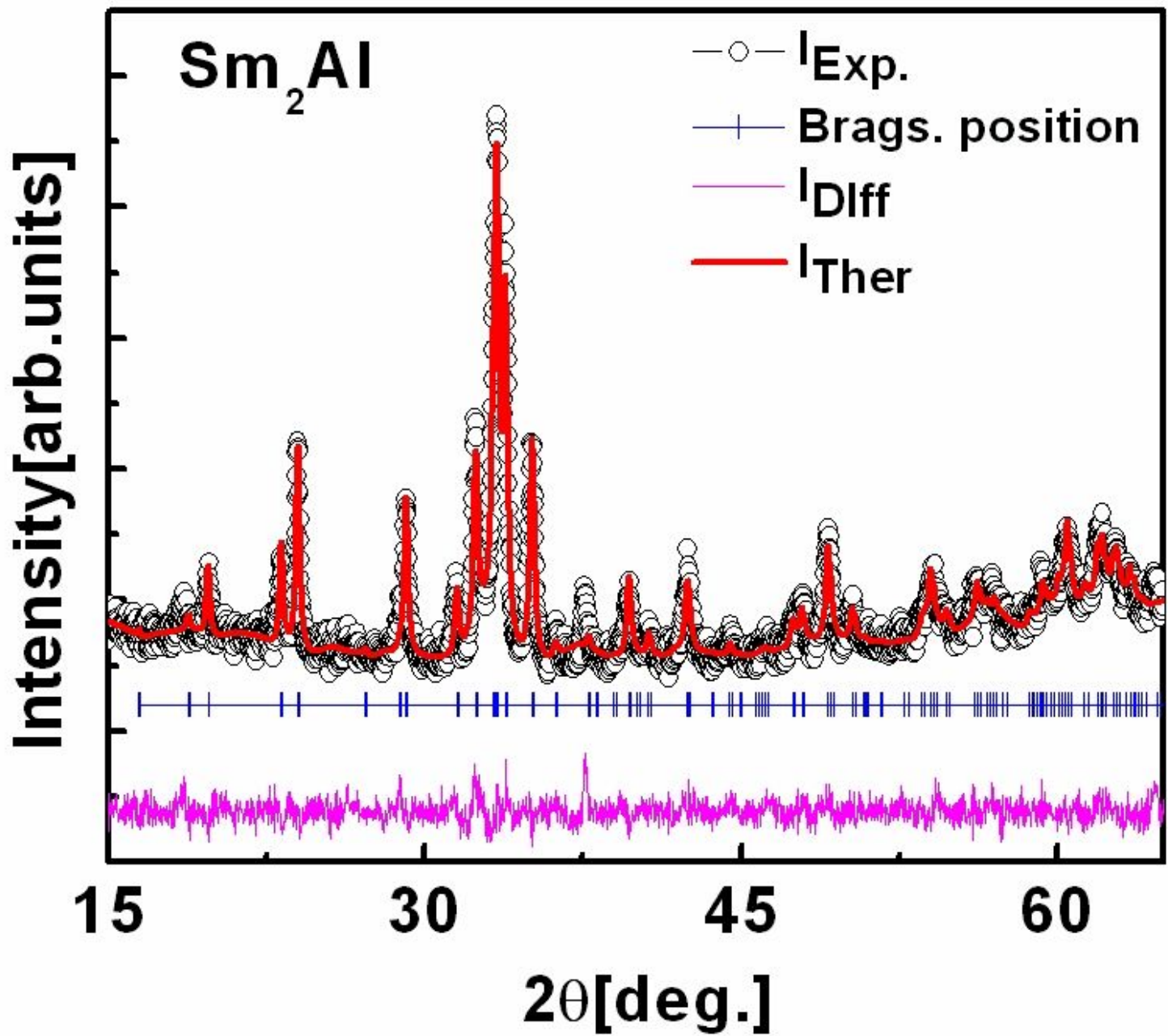

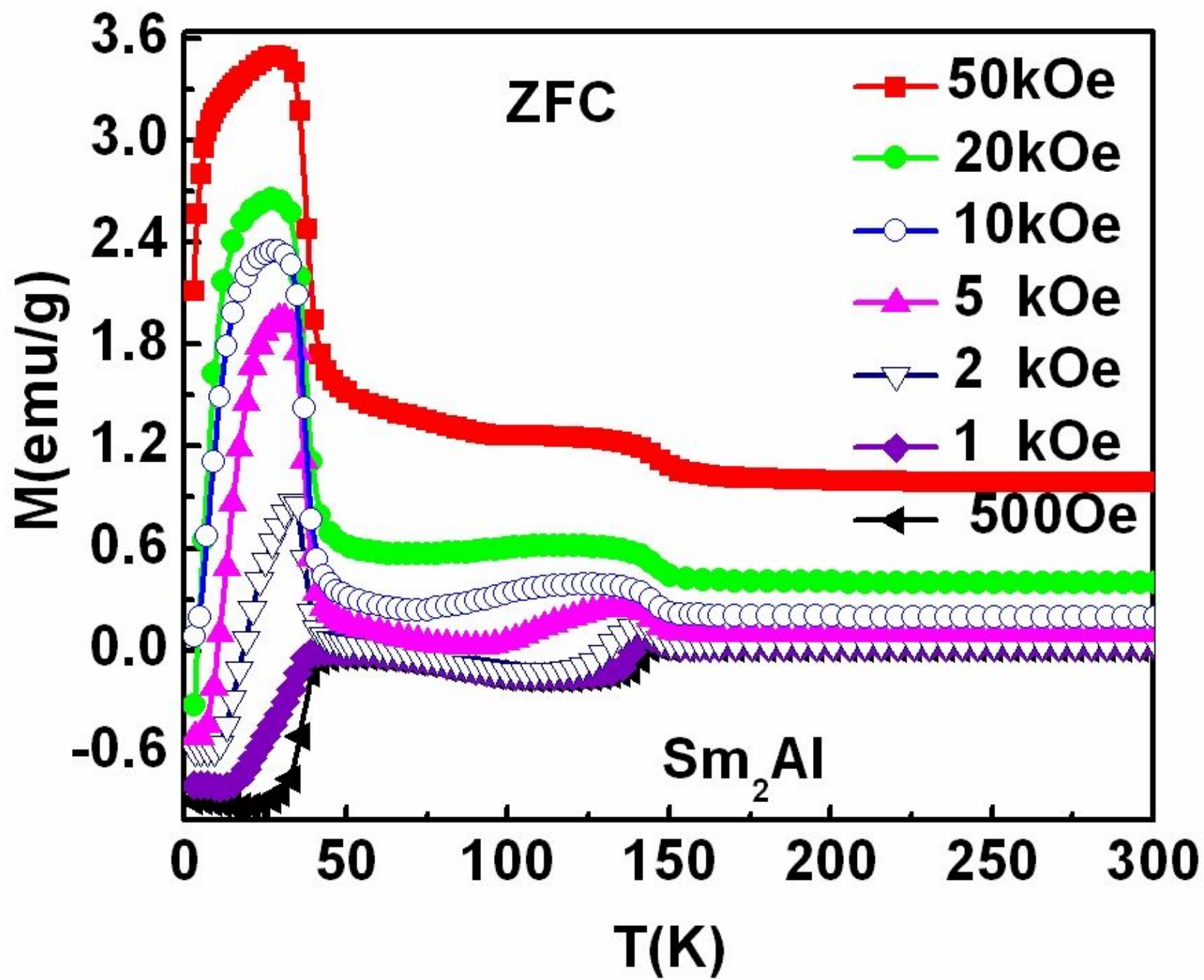

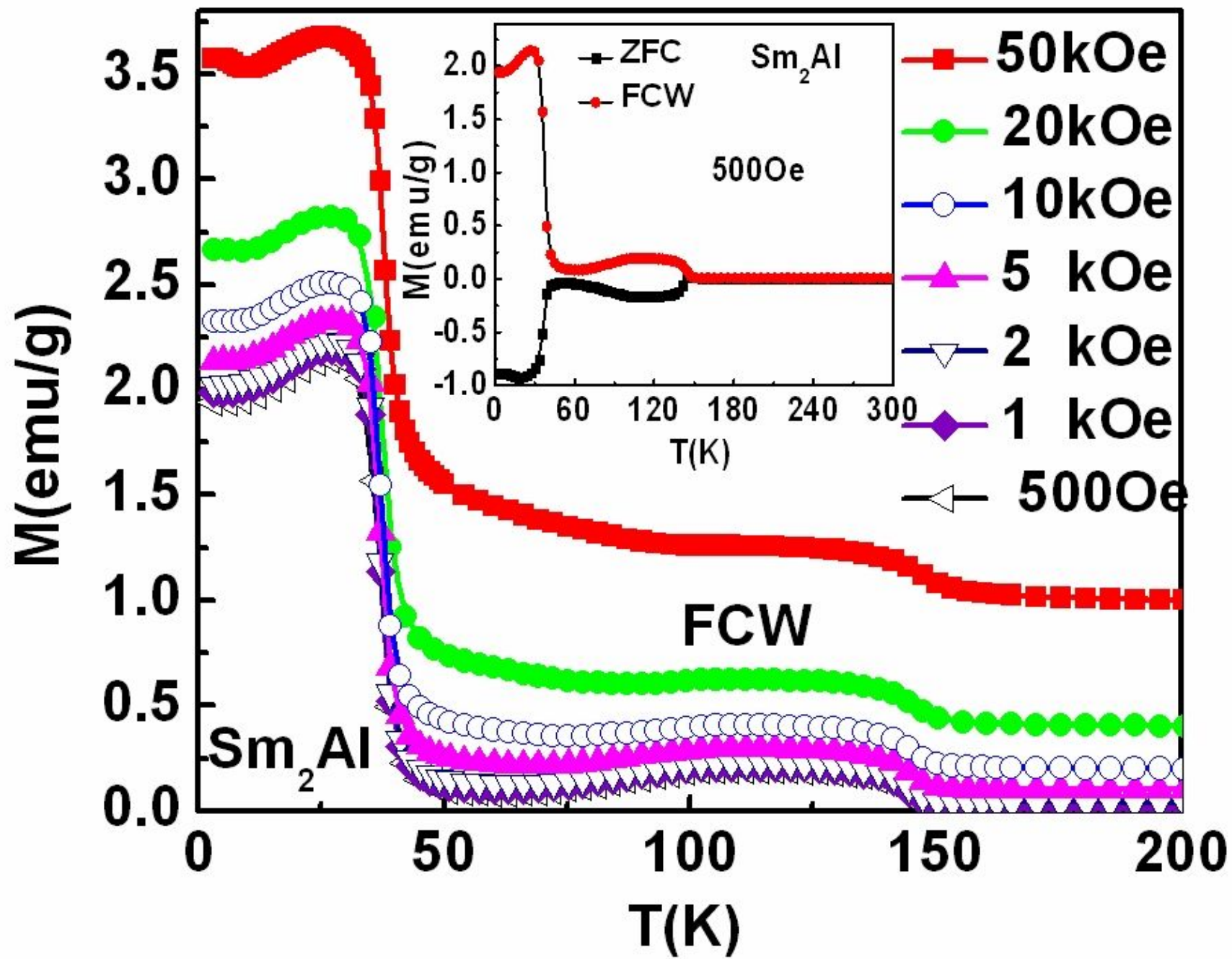

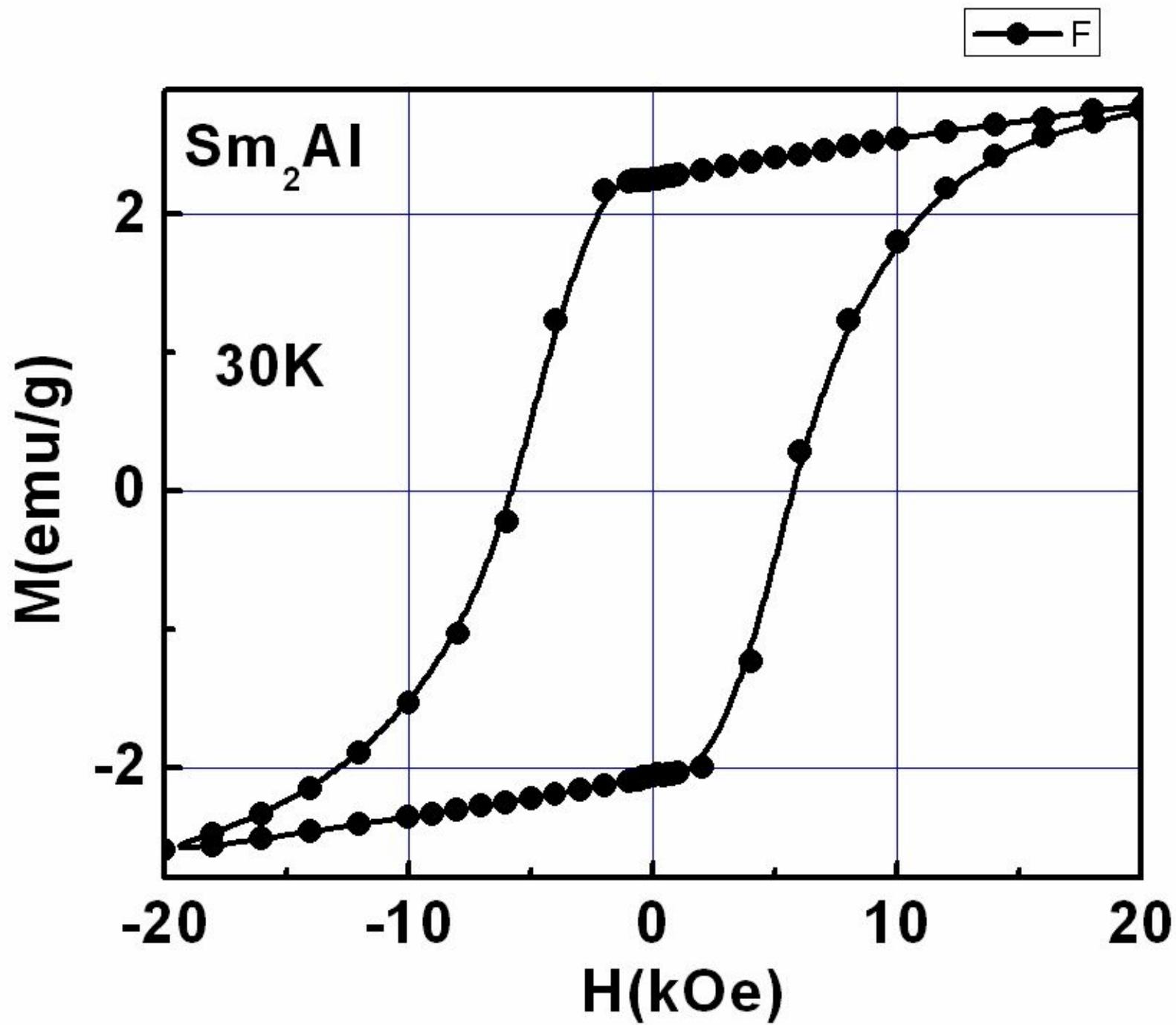

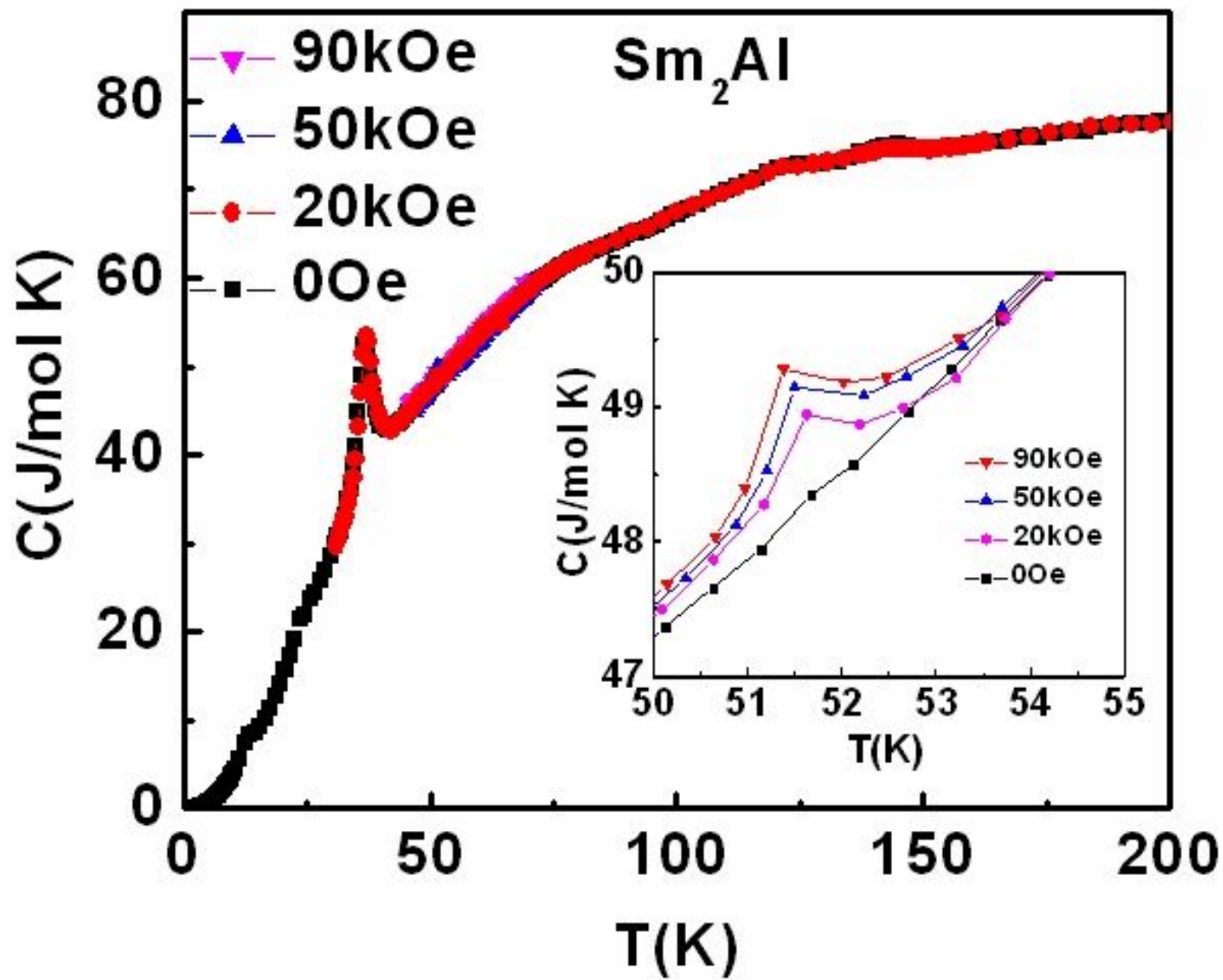